\documentclass[runningheads,a4paper]{llncs}

\usepackage[utf8]{inputenc}
\usepackage[english]{babel}
\usepackage{amssymb}
\usepackage{graphicx}

\usepackage{url}
  
\newcommand{\keywords}[1]{\par\addvspace\baselineskip
\noindent\keywordname\enspace\ignorespaces#1}

\begin{document}

\mainmatter  

\title{Intra-Team Strategies for Teams Negotiating Against Competitor, Matchers, and Conceders}

\titlerunning{Intra-Team Strategies Against Competitor, Matchers, and Conceders}

%
%
\author{Victor Sanchez-Anguix\inst{1} \and Reyhan Aydo\u{g}an\inst{2} \and Vicente Julian\inst{1} \and Catholijn M. Jonker\inst{2}}
\authorrunning{V. Sanchez-Anguix \and R. Aydo\u{g}an \and V. Julian \and C. M. Jonker}

\institute{Universitat Politècnica de València,\\ Departamento de Sistemas Informáticos y Computación \\
Cami de Vera s/n, 46022, Valencia, Spain\\ \email{\{sanguix,vinglada\}@dsic.upv.es} \and
Interactive Intelligence Group\\
Delft University of Technology \\
Delft, The Netherlands\\ \email{\{R.Aydogan,C.M.Jonker\}@tudelft.nl} }

%
%

\toctitle{Lecture Notes in Computer Science}
\tocauthor{Authors' Instructions}
\maketitle

\begin{abstract}
Under some circumstances, a group of individuals may need to negotiate together as a \emph{negotiation team} against another party. Unlike bilateral negotiation between two individuals, this type of negotiations entails to adopt an intra-team strategy for negotiation teams in order to make team decisions and accordingly negotiate with the opponent. It is crucial to be able to negotiate successfully with heterogeneous opponents since opponents' negotiation strategy and behavior may vary in an open environment. While one opponent might collaborate and concede over time, another may not be inclined to concede. This paper analyzes the performance of recently proposed intra-team strategies for negotiation teams against different categories of opponents: \emph{competitors}, \emph{matchers}, and \emph{conceders}. Furthermore, it provides an extension of the negotiation tool G\textsc{enius} for negotiation teams in bilateral settings. Consequently, this work facilitates research in negotiation teams.
\keywords{Negotiation Teams, Collective Decision Making, Agreement Technologies}
\end{abstract}

\section{Introduction}

A negotiation team is a group of two or more interdependent individuals that join together as a single negotiation party because they share some common goals related to the negotiation at hand \cite{thompson96,thompson01}. This kind of party participates in many real life situations like the negotiation between a married couple and a house seller, the negotiation between a group of traveling friends and a booking agency, and the negotiation between two or more organizations. Despite acting as a single party, most of the time negotiation teams cannot be considered as a unitary player. As a matter of fact, team members may have different and conflicting preferences that need to be conciliated when making a team decision regarding the negotiation. Agent-based negotiation teams (ABNT) constitutes a novel topic of research in automated negotiation, where efforts in the last few years focused mostly on bilateral and multiparty negotiations with unitary players \cite{faratin02,lai08,ito10}. Mechanisms that allow ABNT to take decisions on the negotiation process, namely intra-team strategies or team dynamics \cite{sanchez-anguix11,sanchez-anguix12}, are needed in order to support multi-agent systems for complex applications like group travel markets, group buying in electronic commerce, and negotiations between agent organizations (e.g., organizational merging). An intra-team strategy for a specific negotiation protocol (e.g., alternating bilateral negotiation protocol) defines \textit{what} decisions are taken by the negotiation team, and \textit{how} and \textit{when} those decisions are taken.

Although there are some studies investigating negotiation among team members \cite{tambe99}, automated negotiation between a team and an opponent is open to research.  Sanchez-Anguix {\it et al.} have proposed several intra-team strategies \cite{sanchez-anguix11,sanchez-anguix12} for ABNT following the alternating-offers protocol in a bilateral setting. The proposed intra-team strategies have been studied under different environmental conditions to assess the most appropriate intra-team strategy with respect to the given environmental setting \cite{sanchez-anguix11}. However, several assumptions regarding the opponent exist. For instance, it is assumed that the opponent employs a time-based concession tactics such as Boulware or Conceder\cite{faratin98} in a cooperative context. Nevertheless, these assumptions might become inconsistent with some opponents in open and dynamic environment. For instance, an opponent may adopt a strategy like ``take it or leave it'' while another opponent may choose to observe other negotiating agent's behavior and concede accordingly. An immediate question is which intra-team strategies will negotiate well against other types of opponents different than those using time-based tactics.

Without a doubt, an opponent's negotiation attitude may affect on the performance of intra-team strategies. Opponent's behavior is not limited to classic time-based concession strategies. Baarslag {\it et al.} classify the negotiation strategies according to their negotiation behavior against the opponent into four categories~\cite{baarslag11}. These are inverters, conceders, competitors and matchers. Conceders always concede regardless of the opponent's strategy, while competitors do not yield independently of the behavior shown by opponents. A matcher mimics its opponent's behavior while inverter inverts it. When the opponent concedes, the matcher would concede accordingly while the inverter would not. Based on this classification, we investigate how intra-team strategies proposed in the literature perform against opponents belonging to different families of negotiation strategies. To do this, we extend G\textsc{enius} \cite{Lin12} to allow negotiation teams and enable it to perform bilateral negotiations between a team (a group of agents) and an individual agent. The contributions of this paper do not solely focus on the study of intra-team strategies' performance against different types of opponent, but we also describe how G\textsc{enius} has been modified to support such negotiations. This extension will allow researchers to (i) design and test domain independent intra-team strategies, which is desirable given the increasing number of application domains for automated negotiation; (ii) engage negotiation teams  in open environments where any kind of opponent behavior is possible; (iii) make use of a wide repository of negotiation domains, utility functions, and automated negotiators; (iv) focus on the design of intra-team strategies, while leaving simulation aspects to be governed by G\textsc{enius}.

Our contributions are two-folds. First, we extend G\textsc{enius} to support ABNT; thus G\textsc{enius} can facilitate research on ABNT. Second, we analyze the performance of different intra-team strategies proposed by Sanchez-Anguix {\it et al.} against different types of heterogeneous opponents. The rest of this paper is organized as follows. First, we present our general framework. After that, we briefly introduce the intra-team strategies analyzed in this paper. Then, we describe how the extension has been included inside the G\textsc{enius} framework.  Then, we describe how the experiments were carried out and present and discuss the results of the experiments. Finally, we describe our future work and briefly conclude this work.

\section{General Framework}
\label{Sec-NegotiationSetting}

\begin{figure}[t]
\begin{center}
\includegraphics[width=\linewidth]{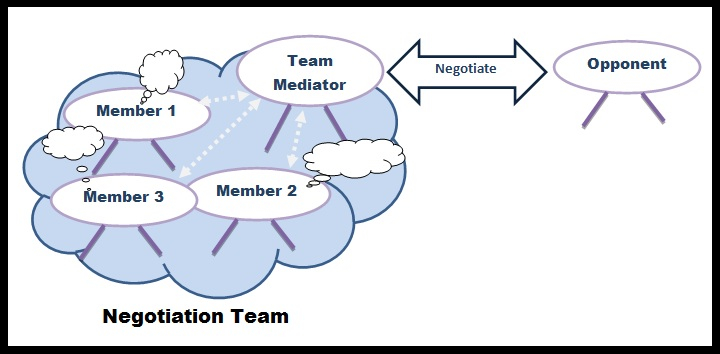}
\caption{This picture shows our general negotiation framework.}
\label{nego}
\end{center}
\end{figure}

In our framework, one negotiation team is involved in a negotiation with an opponent. Independently of whether or not the other party is also a team, both parties interact with each other by means of the alternating-offers protocol.  Team dynamics or intra-team strategies define \textit{what} decisions have to be taken by a negotiation team, \textit{how} those decisions are taken, and \textit{when} those decisions are taken. In a bilateral negotiation between a team and an opponent, the decisions that must be taken are which offers are sent to the opponent, and whether or not opponent's offers are accepted. A general view of our framework is represented in Fig. \ref{nego}. Dashed lines depict communications inside the team, while others represent communications with the opponent.

 A team $A$ is formed by a team mediator $TM_{A}$ and team members $a_{i}$. The team mediator communicates with the other party following the alternating offers protocol and team members communicate with the team mediator. Communications between the team and the opponent are carried out by means of the team mediator. This mediator sends team decisions to the opponent and receives, and later broadcasts, decisions from the opponent to team members. Thus, the fact that the opponent is communicating with a team is not known by the opponent, which only interacts with the trusted mediator. In this framework, the mechanisms employed by the team to decide on which offers to send and whether or not accept offers are carried out during the negotiation process itself. How these decisions are taken depends on the specific intra-team strategy that is implemented by the team and the team mediator. Each team mediator can implement its own intra-team protocol to coordinate team members as long as team members know how to play such intra-team protocol.
It should be noted that we assume that team membership remains static during the negotiation process. Thus, members do not leave/enter the team as the negotiation is being carried out. It is acknowledged that team members may leave or join the group in certain specific situations. However, membership dynamics is not considered in this article, and it is designated as future work.

\begin{figure}[t]
 \centering
 \includegraphics[width=\linewidth]{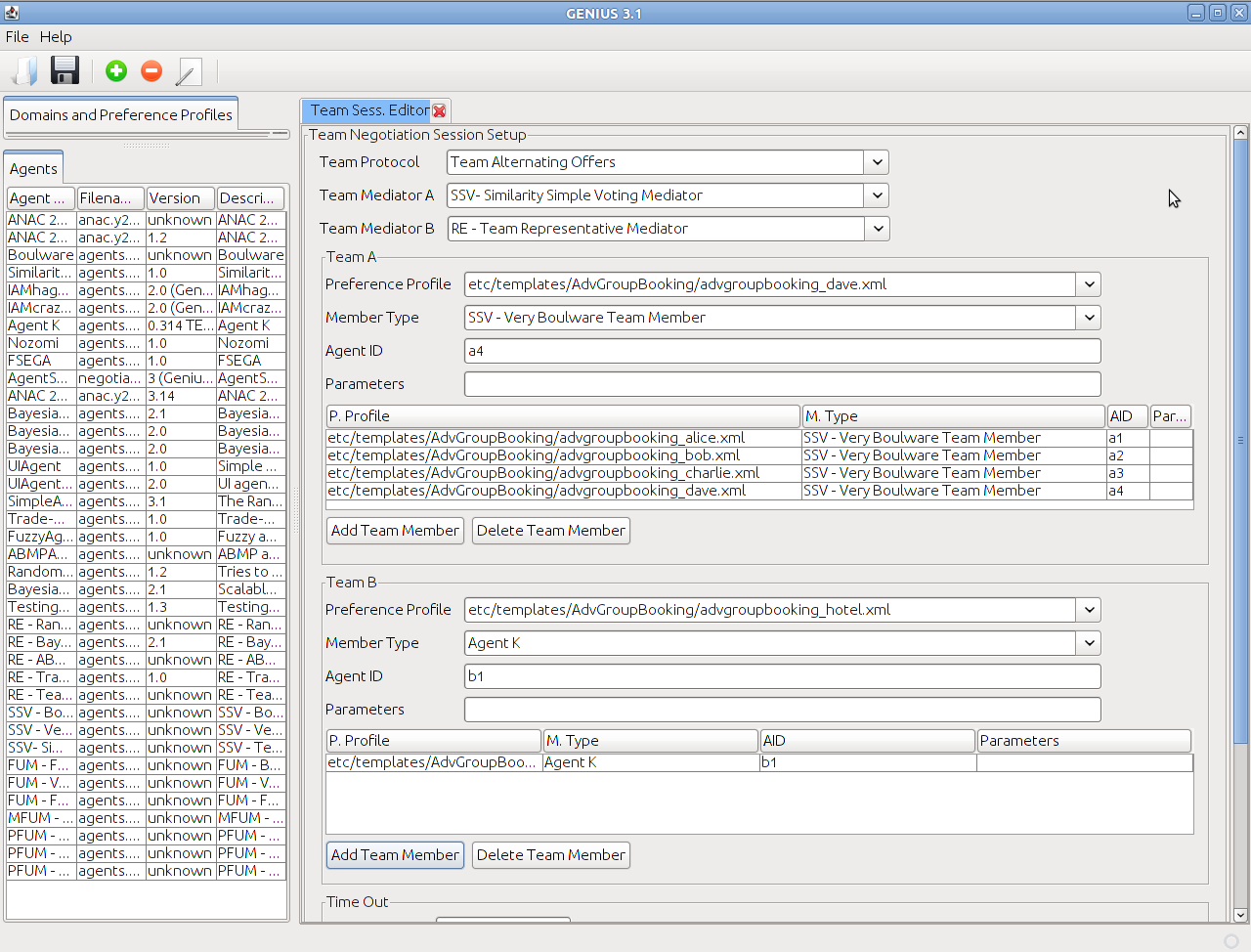}
\caption{Screenshot showing the menu for configuring a team negotiation session.}
\label{fig:genius1}
\end{figure}

\section{Intra-team strategies}
\label{sec-intraTeamStrategies}
In this section we briefly describe the intra-team strategies proposed by Sanchez-Anguix \textit{et al.} \cite{sanchez-anguix11,sanchez-anguix12} that are the focus of our study. These intra-team strategies have been selected according to the minimum level of unanimity that they are able to guarantee regarding each team decision: no unanimity guaranteed (representative), majority/plurality (similarity simple voting), semi-unanimity (similarity Borda voting), and unanimity (full unanimity mediated).
\subsection{Representative (RE)}
The intra-team protocol employed by the representative intra-team strategy is the simplest possible strategy for a negotiation team. Basically, one of the team members is selected as representative and acts on behalf of the team. Interactions among team members are non-existent, and therefore, every decision is taken by the representative according to its own criterion. Obviously, the performance of the team will be determined by the similarity among the team members' utility functions and the negotiation skills of the representative. It is expected that if team members' utility functions are very similar and the representative negotiation strategy is appropriate, the team performance will be reasonably good.

The mechanism used by the team to select its representative may vary depending on the domain: trust, past experiences, rational voting based on who is more similar to oneself, etc. Since G\textsc{enius} is a general simulation framework, a random team member is selected as representative. This random team member will receive the messages from the opponent and act accordingly by sending an offer/counter-offer or accepting the opponent's offer. Generally, any G\textsc{enius} agent that knows how to play the alternating bilateral game can act as representative.

 \subsection{Similarity Simple Voting (SSV)}
 The similarity simple voting intra-team strategy relies on voting processes to decide on which offer is proposed and whether or not the opponent's offer is accepted. Being based on voting, the strategy requires the action of team members and team coordination by means of a mediator. The intra-team strategy goes as follows every round:
 \begin{itemize}
  \item Accept/Reject opponents' offer: The team mediator receives an offer from the other party. Then, the mediator broadcasts this offer to the team members, indicating that it comes from the opponent party. The team mediator opens a voting process, where each team member should respond to the mediator with an Accept or Reject depending on the acceptability of the offer from the point of view of the team member. The team mediator gathers the responses from every team member and applies a majority rule. If the number of Accept actions received from team members is greater than half the size of the team, the offer is accepted and the corresponding Accept action is sent to the opponent. Otherwise, the team mediator starts the offer proposal mechanism.
  \item Offer Proposal: Each team member is allowed to propose an offer to be sent. This offer is communicated solely to the mediator, who will make public the offers, and start a voting process. In this voting process, each team member must state whether or not he considers it acceptable for each of the offers proposed. For instance, if three different offers $(x_{1},x_{2},x_{3})$ have been proposed, the team members should state about the acceptability of the three of them: e.g., $(yes,no,yes)$. The mediator applies a plurality rule to determine the most supported offer, which is the one that is sent to the opponent.
 \end{itemize}

 Standard team members for SSV employ time tactics to decide on the acceptance/rejection of the opponent's offer, and the offer proposal. More specifically, the current aspiration for team members follows the next expression \cite{lai08}:
\begin{equation}
  s_{a_{i}}\left(t\right)=1-\left(1-RU_{a_{i}}\right)\left(\frac{t}{T_{A}}\right)^{1/\beta_{a_{i}}}\enspace .
 \label{eq:time}
 \end{equation}
  where $s_{a_{i}}(t)$ is the utility demanded by the team member at time $t$, $RU_{a_{i}}$ is the reservation utility for the agent, $T_{A}$ is the team's deadline, and $\beta_{a_{i}}$ is the concession speed. When $\beta_{a_{i}}<1$ we have a classic Boulware strategy, when $\beta_{a_{i}}=1$ we have a linear concession, and when $\beta_{a_{i}}>1$ we have a conceder strategy. On the one hand a team members considers an opponent's offer as acceptable if it reports a utility which is greater than or equal to $s_{a_{i}}(t)$. On the other hand, a team member considers an offer proposed by a teammate acceptable if it reports a utility which is greater than or equal to the utility of the offer that he proposed to the team in the same round. As for the offer proposed to the team, team members attempt to select the offer from the iso-utility curve which is closer to the last opponent's offer and the offer sent by the team in the last round (similarity heuristic based on Euclidean distance).

\subsection{Similarity Borda Voting (SBV)}
This intra-team strategy attempts to guarantee a higher level of unanimity by incorporating voting mechanisms that select broadly accepted candidates like Borda count \cite{nurmi10}, and unanimity voting processes. The intra-team strategy proceeds as SSV with the following differences:
\begin{itemize}
 \item Accept/Reject opponents' offer: Instead of using a majority voting to decide whether or not the opponent's offer is accepted, the team mediator opens a unanimity voting. Hence, an opponent offer is accepted if and only if the number of Accept actions is equal to the number of team members. Otherwise, the team mediator stars the offer proposal mechanism.
\item Offer Proposal: Each team member is allowed to propose an offer to be sent by the same mechanisms described in SSV. Then, the team mediator makes the team members privately score each proposal by means of a Borda count. The team members give a different score to each offer from the set $[0,|A|-1]$, where $|A|$ is the number of proposals received by the team mediator. Once the scores have been received by the team mediator, it selects the candidate offer that received the highest sum of scores and it is sent to the opponent.
\end{itemize}

Standard team members are governed by an individual time-based concession tactic like the one in Equation \ref{eq:time}. Similarly to SSV, an opponent's offer is acceptable for a team member if the utility that it reports is equal to or greater than its current utility demanded $s_{a_{i}}(t)$. When scoring team offers, each team member privately ranks candidates in descending order of utility and then it assigns a score to each offer which is equal to the number of candidates minus the position of the offer in the ranking.

\subsection{Full Unanimity Mediated (FUM)}
FUM is capable of reaching unanimous decisions as long as the negotiation domain is composed by predictable issues whose type of valuation function is the same for team members (e.g., either monotonically increasing or decreasing). The type of unanimity that it is capable of guaranteeing is strict in the sense that every decision reports a utility which is greater than or equal to the current aspiration level of each team member \cite{sanchez-anguix12}. The team mediator governs intra-team interactions as follows:
\begin{itemize}
 \item Accept/Reject opponents' offer: The interaction protocol followed by the team in this decision is the same as the one presented in SSV. However, the decision rule applied by the team mediator in this case is unanimity. Therefore, an opponent offer is only accepted if it is acceptable to each team member.
 \item Offer Proposal: Every team member is involved in the offer proposal, which consists in an iterated process where the offer is built attribute per attribute. The mediator starts the iterated building process with an empty partial offer (no attribute is set). Then, he selects the first attribute to be set following an agenda. The mediator makes public the current partial offer and the attribute that needs to be set. Each active team member states privately to the mediator the value that he wants for the requested issue. When all of the responses have been gathered, the mediator aggregates the values sent by team members using the max (monotonically increasing valuation function) or min (monotonically decreasing valuation function) and makes public among active team members the new partial offer. Since it is assumed that team members share the same type of valuation function for predictable attributes, increasing the welfare of one of the members results in other team members increasing their welfare or staying at the same utility. Then, each active team member must evaluate the partial offer and state if the partial offer is acceptable at the current state (Accept or Reject action). Those team members that respond with an Accept action are no longer considered active in the current construction process. The team mediator selects the next attribute in the agenda and follows the same process until all of the attributes have been set or until there are no more active team members (the rest of the attributes are maximized to match the opponent's preferences). It should be said that the agenda of attributes is set by the mediator observing the concessions from the opponent in the first interactions. Following a rational criteria, the opponent should have conceded less in the most important attributes in the first negotiation rounds. The amount of concession in each attribute during the first rounds is summed up and an attribute agenda is inferred at each round. The first attributes in the agenda are those inferred as less important for the opponent (more amount of concession), whereas the last attributes in the agenda are those considered more important for the opponent. The heuristic behind this agenda and iterated building process is attempting to satisfy team members first with those attributes less important for the opponent.
\end{itemize}

As for the standard team member behavior, team members have their demands governed by an individual time-based concession tactic like the one in Equation \ref{eq:time}. In the iterated building process, each team member requests the attribute value which, given the current partial offer, makes the partial offer closer to its current demands $s_{a_{i}}(t)$. Additionally, a partial offer is acceptable when, considering only those attributes that have been set, the partial offer reports a partial utility which is greater than or equal to the current demands of the team member. Each team members considers an opponent offer acceptable when it has a utility which is greater or equal than its current demands $s_{a_{i}}(t)$.

\section{Implementation in Genius}

G\textsc{enius} \cite{Lin12} is a well-known negotiation simulation framework. It supports simulation of sessions and tournaments based on bilateral negotiations. Users are able to design their own agents and test them against a wide variety of different agents designed by the community. The framework provides information critical for analysis (e.g., utility, Pareto optimality, etc.) which is extremely useful for research tasks. Moreover, the use of G\textsc{enius} as a testbed for bilateral negotiations is testified by its use in the annual automated negotiating agent competition (ANAC) \cite{baarslag10}. The ANAC competition provided G\textsc{enius} with a large repository of agents. The repository of available agents contains conceder, inverter, matcher, and competitor agents. The integration of ABNT in G\textsc{enius} additionally facilitates the following objectives:
\begin{itemize}
 \item The framework includes several negotiation domains and utility functions for test purposes. Even though most of these domains are thought for bilateral negotiations with unitary players, it is possible to add new negotiation domains and utility functions in an easy way. In fact, we are in the process of adding new team negotiation domains (i.e., advanced hotel group booking) besides the one employed for the experiments of this paper (i.e., hotel group booking, see Section \ref{sec:domain}).
 \item The use of G\textsc{enius} in ANAC has provided with wide variety of conceder, matcher, inverter and competitor opponents. Previous research in ABNT had only considered opponents with time-based tactics \cite{sanchez-anguix11}.
 \item Current research in ABNT has only considered team members following the same kind of homogeneous behavior inside the intra-team strategy, which may not be the case in some open environments. Due to its open nature, G\textsc{enius} may be able to simulate ABNT whose team members are heterogeneous since they have been designed by different scholars.
 \item G\textsc{enius} is a consolidated testbed among the agent community. Thus, the inclusion of ABNT inside G\textsc{enius} can facilitate research on ABNT by other scholars, and even give room to a future negotiating competition involving teams.
 \item Researchers can either design new team members for teams following the intra-team protocols included in the framework (i.e., team mediators), or they can design new intra-team protocols and team members.
\end{itemize}

In order to implement negotiation teams, two new classes have been introduced in G\textsc{enius}: \textit{TeamMediator} and \textit{TeamMember}. These two classes can be extended by users to include new intra-team strategies and types of team members in the system. Next, we depict the main traits of these classes, and how they can be used to include new features in G\textsc{enius}:
\begin{itemize}
 \item \textit{TeamMember}: Team members extend the \textit{Agent} class, so they have all of their methods available. Actions that come from the opponent party are received by the \textit{ReceiveMessage} method, whereas actions that come from the team mediator are received in the \\ \textit{ReceiveTeamMessage} method. The method \textit{chooseAction} is used to decide the agent's action independently of whether or not the next action involves communications with the team mediator or the opponent.
 \item \textit{TeamMediator}: The team mediator is the agent that communicates with the opponent party, and transmits opponent's decisions to the team members. Thus, it has access to the public interface of all of the team members. Depending on the kind of intra-team strategy, the mediator also coordinates other processes like voting mechanisms, offer proposal mechanisms, and so forth. As the \textit{Agent} class, it receives communications from opponent by the \textit{ReceiveMessage} method. In the \textit{chooseAction} method, the mediator can either directly send a decision to the opponent, or communicate with team members to decide the next action to be taken. When interacting with team members, the mediator uses the \textit{ReceiveTeamMessage} method in team members' API to send messages to team members. Team members can respond to the mediator with the \textit{chooseAction} method in their public API. The TeamMediator class is completely flexible, as the only mandatory action is receiving opponent's decisions and sending decisions to the other party. Therefore, any kind of mediated communication protocol can be implemented extending the \textit{TeamMediator} class.
\end{itemize}

Of course, it should be noted that team members and team mediators are tightly coupled. For a team member to participate in a negotiation team governed by a specific mediator, the team members should know the intra-team communication protocol implemented by such mediator.

G\textsc{enius} provides several measures to assess the quality of negotiating agents. The current version of G\textsc{enius} is capable of running team negotiation sessions between two parties and provide online information about the minimum utility of team members, the average utility of team members, the maximum utility of team members, the joint utility of team members, current round, and current negotiation time for each offer exchanged between both parties. A screenshot of the environment being configured for a team negotiation session can be observed in Fig. \ref{fig:genius1}. In the upper part of the menu, the user can select the intra-team strategies to be used by each party, whereas the user add and remove team members for each party in the lower part of the menu.

\section{Experiments and Results}
As stated in the Section 1, one of the purposes of this paper is assessing the performance of intra-team strategies against negotiation strategies different from classic time-based tactics. With that purpose, we tested RE, SSV, SBV and FUM against agents from the ANAC 2010 competition who have been previously classified into competitors, conceders, and matchers\footnote{Because of the technical inconsistencies, we could not use ANAC's inverters directly in our settings. Thus, they are not included in this analysis.} \cite{baarslag11}. First, we briefly describe the agents that we selected from the agent competition to represent the different families of negotiation strategies. Then, we introduce the negotiation domain used for the experiments. After that, we describe how the experiments were carried out, and, finally, we show and analyze the results of the experiments.
\subsection{ANAC 2010 Agents}
\label{sec-ANACAgents}
In this section we present the different ANAC 2010 agents employed for our experiments. These agents pertain to three of the four categories presented previously: matchers, conceders, competitors.
\begin{itemize}
 \item IAMHaggler \& IAMCrazzyHaggler \cite{williams10,baarslag10}: On the one hand, IAMCrazyHaggler is basically a take-or-leave it agent that proposes offers over a high threshold. The only aspect taken into consideration for accepting an offer is the utility of such offer, and not time. Due to this behavior, the experiments carried out in \cite{baarslag11} classified IAMCrazyHaggler as the most competitive strategy in the ANAC 2010 competition.  On the other hand, IAMHaggler is a much more complicated agent. It employs Bayesian learning and non-linear regression to attempt to model the opponent party and updates its acceptance threshold based on information like time, the model of the opponent, and so forth. It was classified as a competitor agent in the experiments carried out in \cite{baarslag11}
 \item Agent Smith \cite{galen12,baarslag10}: This agent is a conceder agent \cite{baarslag11} that starts by demanding the highest utility for himself and slowly concedes to attempt to satisfy the preferences of the opponent by means of a learning heuristic. When the timeline is approaching two minutes, it proposes the best offer received up until that moment in an effort to finish the negotiation.
 \item Agent K \cite{kawaguchi11,baarslag10}: This agent was the winner of the ANAC 2010 competition.  It adjusts its aspirations (i.e., target utility) in the negotiation process considering an estimation of the maximum utility that will be offered by the other party. More specifically, the agent gradually reduces its target utility based on the average utility offered by the opponent and its standard deviation. If an offer has been proposed by the opponent that satisfies such threshold, it is sent back since, rationally, it should be also good enough for the opponent. In \cite{baarslag11}, Agent K was classified as a competitor agent.
 \item Nice Tit-for-Tat \cite{hindriks09,baarslag11,baarslag13}: This strategy  is a matcher agent from the 2011 ANAC competition that reciprocates the other party's moves by means of a Bayesian model of the other party's preferences. According to the Bayesian model, the Nice Tit-for-Tat agent attempts to calculate the Nash point and it reciprocates moves by calculating the distance of the last opponent offer to the aforementioned point. When the negotiation time is reaching its deadline, the Nice TFT agent will wait for an offer that is not expected to improve in the remaining time and accept it in order to secure an agreement.
 \end{itemize}

\subsection{Test Domain: Hotel Group Booking}
\label{sec:domain}
A group of friends who have decided to spend their holidays together has to book accommodation for their stay. Their destination is Rome, and they want to spend a whole week. The group of agents engages in a negotiation with a well-known hotel in their city of destination. Both parties have to negotiate the following issues:
 \begin{itemize}
  \item Price per person ($pp$): The price per person to pay.
  \item Cancellation fee per person ($cf$): The fee that should be paid in case that the reservation is cancelled.
  \item Full payment deadline ($pd$): It indicates when the group of friends has to pay the booking.
  \item Discount in bar ($db$): As a token of respect for good clients, the hotel offers nice discounts at the hotel bar.
 \end{itemize}
In our experimental setup, preference profiles are represented by means of additive utility functions in the form:
\begin{equation}
  U_{p_{i}}\left(X\right)=w_{p_{i},1} V_{p_{i},1}\left(x_{1}\right) + ... + w_{p_{i},n} V_{p_{i},n}\left(x_{n}\right)\enspace .
 \end{equation}
 where $w_{p_{i},j}$ is the weight given by agent $p_{i}$ to attribute $j$, $V_{p_{i},j}$ is the valuation function for attribute $j$, and $x_{j}$ is the value of attribute $j$ in the offer $X$. The domain of the attribute values is continuous and scaled to [0,1]. It should be noted that all of the team members share the same type of monotonic valuation function for the attributes (monotonically increasing for payment deadline and discount, and decreasing for price and cancellation fee) so that there is potential for cooperation among team members. Despite this, team members give different weights to the negotiation issues. The type of valuation function for the opponent is the opposite type (increasing for price and cancellation fee, and decreasing for the payment deadline and the discount) and weights may be different too. The preference profiles of the agents can be found in Table \ref{tab:preference}. Even though SSV, SBV, and RE are able to handle other types of domain where unpredictable attributes are present, we only use domains with predictable attributes in our analysis because FUM does not support domains having unpredictable attributes.

\begin{table}[t]
\center

 \begin{tabular}{l| l l l l}
 & $w_{pd}$ & $w_{cf}$ & $w_{pd}$ & $w_{db}$ \\
\hline
$a_{1}$& 0.5 & 0.1 & 0.05 & 0.35 \\
\hline
$a_{2}$& 0.25 & 0.25 & 0.25 & 0.25 \\
\hline
$a_{3}$& 0.30 & 0.50 & 0.05 & 0.15 \\
\hline
$op$&    0.10 & 0.50 & 0.25 & 0.15 \\
\hline
 \end{tabular}
 \caption{Preference profiles used in the experiments. $a_{i}$ represents team members and $op$ represents the opponent.}
\label{tab:preference}
\end{table}
\subsection{Experimental Setting}

In order to evaluate the performance of the intra-team strategies introduced in Section~\ref{sec-intraTeamStrategies}, we set up a negotiation team consisting of three members that negotiates with each ANAC agent presented in Section~\ref{sec-ANACAgents}. We tested the intra-team strategies with different parameters' configurations: the FUM strategy where the concession speed of each team member is drawn from the uniform distribution $\beta_{a_{i}}=U[0.5,0.99]$ (FUM Boulware or FUM B) or $\beta_{a_{i}}=U[0.01,0.4]$ (FUM Very Boulware or FUM VB), the SSV strategy where $\beta_{a_{i}}=U[0.5,0.99]$ (SSV Boulware or SSV B) or $\beta_{a_{i}}=U[0.01,0.4]$ (SSV Very Boulware or VB), the SBV strategy where $\beta_{a_{i}}=U[0.5,0.99]$ (SBV Boulware or SBV B) or $\beta_{a_{i}}=U[0.01,0.4]$ (SBV Very Boulware or VB), and the representative approach employing Agent K as the negotiation strategy (RE K). Both parties have a shared deadline $T=180$ seconds. If the deadline is reached and no final agreement has been found, both parties get a utility equal to 0.

In the experiments, each intra-team strategy was faced against each ANAC agent 10 different times to capture stochastic differences in the results. Out of those 10 repetitions, half of the times the initiating party was the team, and the other half the initiating party was the ANAC agent. We gathered information on the average utility of team members in the final agreement, and the joint utility of both parties (product of utilities of team members and opponent). A one-way ANOVA ($\alpha=0.05$) and a post-hoc analysis with Tukey's test was carried out to assess the differences in the averages.
\subsection{Results}

One of the goals of this paper is identifying which intra-team strategies work better against different opponents. Therefore, we start by analyzing the results for the average utility of team members in Table \ref{tab:mean}. The results in bold fond indicate which intra-team strategy obtains statistically better results according to ANOVA ($\alpha=0.05$) and post-hoc analysis with Tukey's test. As expected, all of the intra-team strategies, especially when their concession speed is very boulware, get higher average utility for team members while negotiating with an opponent employing a conceder strategy like Agent Smith than while negotiating with competitive opponents like AgentK, IAMHaggler, and IAMCrazyHaggler. This result supports the observation of \cite{baarslag11} that a successful negotiating agent, if we only consider a single negotiation with the opponent (short term relationship), should behave competitively, especially against cooperative strategies.

\begin{table}[t]
\center

 \begin{tabular}{|l | l | l | l | l | l |}
\hline
  & \multicolumn{3}{| c |}{Competitive} & Matcher & Conceder \\
\hline
 & Crazy & Haggler & K & TFT & Smith \\
\hline
FUM B&     \textbf{0.19}  &    0.38      &  0.29    &    \textbf{0.72}  &  0.68       \\
\hline
FUM VB&      0.16    &  \textbf{0.42}        &  \textbf{0.65}   &   \textbf{0.72}     &  \textbf{0.97}         \\
\hline
RE K  &       0.00  &    0.26     &    0.57    &   \textbf{0.70}    &  0.86         \\
\hline
SSV B &       0.14  &     0.34    &   0.36  &   0.45    &   0.57      \\
\hline
SSV VB&        0.08    &   0.36     &  0.57   &  0.44    &  \textbf{0.98}         \\
\hline
SBV B &     0.14    &   0.35      &  0.31   &   0.49    &   0.55      \\
\hline
SBV VB&       0.13     &   0.39     &   0.59  &  0.50    &    \textbf{0.98}       \\
\hline
 \end{tabular}

 \caption{The table shows the average of the average utility for team members in the final agreement. Crazy: IAMCrazyHaggler, Haggler: IAMHaggler,K: Agent K, TFT: Nice Tit-for-Tat, Smith: Agent Smith, B: $\beta=U[0.5,0.99]$, VB: $\beta=U[0.01,0.49]$}
\label{tab:mean}
\end{table}

When the opponent is a conceder (Agent Smith),  we observe that, in our experiments, the best intra-team strategies are those that wait as much as possible to concede and exploit the opponent. We refer to FUM, SBV, and SSV strategies employing a very Boulware time tactic (FUM VB, SBV VB, and SSV VB).  FUM VB, SBV VB, and SSV VB statistically get the same average for the average utility of team members\footnote{One-way ANOVA alpha=0.05 and a post-hoc Tukey test was carried out to support our claims.}. This can be explained due to the fact that the conceder agent has fully conceded before FUM, SBV, and SSV VB have started to concede.  Since a concession from the opponent generally results in all of the team members increasing their welfare, these intra-team strategies perform similarly even though they ensure different levels of unanimity regarding team decisions. A representative using agent K also performs reasonably well due to the same reason. However, since only one of the team members takes decisions, it may not reach an average utility comparable to the ones obtained by FUM VB, SBV VB, and SSV VB.

When the opponent is a matcher, it is observed that employing FUM strategies (B and VB) and a representative strategy with a competitor representative (RE K) results in higher average utility for the team than employing SSV and SBV strategies (B and VB). According to the one-way ANOVA test, the performances of FUM strategies (B and VB) and RE K are statistically and significantly better than those of SSV and SBV strategies (B and VB). The fact that SSV and SBV do not guarantee unanimity regarding team decisions has an important impact on the average utility of team members when faced against Nice Tit-for-Tat. There are no significant differences between FUM B, FUM VB, and RE K. Even though using an Agent K representative guarantees less unanimity regarding team decisions than other intra-team strategies, it is shown that, against certain types of opponents, a representative with a competitor negotiation strategy may be enough in practice to achieve results comparable to results obtained by strategies that guarantee unanimity like FUM.

When the opponent is a competitor, team strategies that employ FUM VB, SBV VB, and SSV VB perform better than their correspondents, FUM B, SBV B, and SSV B respectively. That is, if the opponent is competitive, taking a competitive approach and conceding less results in better average team utility than taking cooperative approaches. In any case, we can observe that the average utility obtained by team members in some competitive settings (i.e., against Haggler and IAMCrazyHaggler) is way lower than the one obtained by the same intra-team strategies against conceders or matchers. This suggests the necessity to explore new intra-team strategies that are able to cope with some competitor agents. If we compare the performances of FUM VB, SBV VB, and SSV VB, the results show that the team using FUM VB gathers higher utility on average than the rest of the cases. The fact that FUM approaches are usually the best options may be explained due to the fact that it ensures that all of the team members are satisfied with those offers sent to the opponent and offers sent by the opponent.  Note that when the opponent is IAMCrazyHaggler, which is a take it or leave it strategy, FUM B gets a higher average for the average utility of team members, but there is no statistically and significant difference with the second runner, FUM VB. In any case, FUM B is statistically different that the rest of intra-team strategies. In this case, RE K is not capable of retaining an average utility for team members comparable to FUM VB, SBV VB, and SSV VB. In fact, all of the negotiations between RE K and IAMCrazyHaggler failed.

\begin{table}[t]
\center

 \begin{tabular}{|l | l | l | l | l | l |}
\hline
  & \multicolumn{3}{| c |}{Competitive} & Matcher & Conceder \\
\hline
 & Crazy & Haggler & K & TFT & Smith \\
\hline
FUM B&   \textbf{0.005}    &    0.04 &   0.03   &  \textbf{0.17}   &   \textbf{0.16}     \\
\hline
FUM VB&    0.004      &   \textbf{0.06}  &  \textbf{0.15}    & \textbf{0.17}       &   0.04        \\
\hline
RE K  &   0.00     &   0.05    &  0.11     &   \textbf{0.15}    &  0.09         \\
\hline
SSV B &   0.002     &    0.03     &  0.04  &  0.07     &  0.10       \\
\hline
SSV VB&    0.001       &    0.04   &  0.10   &  0.04    &  0.02        \\
\hline
SBV B &   0.002     &  0.03       &  0.03  &   0.06    &   0.09      \\
\hline
SBV VB&    0.002       &  0.05     &   0.11  &  0.06    &   0.02       \\
\hline
 \end{tabular}

 \caption{The table shows the average for the joint utility (product) of team members and opponent in the final agreement.}
\label{tab:joint}
\end{table}

Our second evaluation metric is the joint utility of the final agreement\footnote{The product of the utilities of each team member and the opponent}. The joint utility of all of the participants is a crucial metric in situations where both parties not only want to get a deal, but also build a long term relationship and engage in multiple negotiations in the future. An agent that has been exploited in the negotiation process may be reluctant to negotiate with the same team/opponent in the future. Table 2 shows the average for the joint utility in the final agreement. The best intra-team strategies for each opponent in the average utility case, are also the best intra-team strategies in the joint utility case. The only exception to this rule is the conceder case. In that scenario, the best results are obtained by employing  FUM with a Boulware strategy  instead of exploiting the opponent with very Boulware strategies (FUM VB, SSV VB). Thus, if a long term relationship is to be built with conceder agents, it may be wise to employ more concessive intra-team strategies. Very Boulware strategies exploit the opponent, and get very high results for the average utility of team members, but they do not allow the opponent to get high utilities, which results in low joint utilities. On average the highest joint utility is gathered when the team employs a FUM strategy against to a matcher opponent namely, Nice Tit for Tat. Since a matcher matches its opponent, TFT matches its behavior with FUM. Even if FUM concedes slowly over time, TFT will also concede, precluding both parties from being exploited. In competitive settings, FUM needs to adjust its concession speed (very Boulware for IAMHaggler and Agent K, and Boulware for IAMCrazyHaggler)  to be able to get the most for the joint utility, and, still, the results are specially low against Hagglers. This again suggests the necessity to explore new intra-team strategies that are able to cope with some competitor agents.

\section{Related Work}
\label{Sec-Related}

The artificial intelligence community has focused on bilateral or multi-party negotiations where parties are composed of single individuals. The most relevant difference in our work is that we consider multi-individual parties. Next, we analyze and discuss work related in artificial intelligence. 

First, we review some relevant work in bilateral negotiations. Faratin \emph{et al.} \cite{faratin98} introduced some of the most widely used families of concession tactics in negotiation. The authors proposed concession strategies for negotiation issues that are a mix of different families of concession tactics. The authors divide these concession tactics into three different families: (i) time-dependent concession tactics; (ii) behavior-dependent concession tactics; and (iii) resource-dependent tactics. Our negotiation framework also considers time as crucial element in negotiation. Therefore, team members employ time tactics inspired in those introduced by Faratin \emph{et al.}. In another work, Lai \emph{et al.} \cite{lai08} propose an extension of the classic alternating bargaining model where agents are allowed to propose up to $k$ different offers at each negotiation round. Offers are proposed from the current iso-utility curve according to a similarity mechanism that selects the most similar offer to the last offer received from the opponent. This present work also considers extending the bilateral alternating protocol by included layers of intra-team negotiation among team members. This way, team members can decide on the actions that should be taken during the negotiation. Robu \emph{et al.}\cite{robu05,robu08} introduce a bilateral negotiation model where agents represent their preferences by means of utility graphs. Utility graphs are graphical models that represent binary dependencies between issues. The authors propose a negotiation scenario where the buyer's preferences and the seller's preferences are modeled through utility graphs. The seller is the agent that carries out a more thorough exploration of the negotiation space in order to search for agreements where both parties are satisfied. With this purpose, the seller builds a model of the buyer's preferences based on historic information of past deals and expert knowledge about the negotiation domain. Differently to this work, we introduce multi-individual parties and add layers of intra-team negotiation to make it possible for team members to decide on which actions to take during the negotiation.

With regards to multi-party negotiations, several works have been proposed in the literature \cite{ehtamo01,klein03,nguyen04,an06,ito10,williams12,mansour12}. For instance, Ehtamo \emph{et al.} \cite{ehtamo01} propose a mediated multi-party negotiation protocol which looks for joint gains in an iterated way and a single agreement should be found to satisfy all of the parties. The algorithm starts from a tentative agreement and moves in a direction according to what the agents prefer regarding some offers' comparison. Klein et al. \cite{klein03} propose a mediated negotiation model which can be extended to multiple parties negotiating the same agreement. Similarly, Ito \emph{et al.} \cite{ito10} propose different types of n-ary utility functions and efficient multiparty models for multiple parties negotiating on the same agreement. Marsa-Maestre \emph{et al.} \cite{marsa09,marsa09b} carry out further research in the area of negotiation models for complex utility functions. More specifically, they extend the constraint based model proposed by Ito et al. \cite{ito10} by proposing different bidding mechanisms for agents. One-to-many negotiations and many-to-many negotiations also represent special cases of multi-party negotiations. One-to-many negotiations represent settings where one party negotiates simultaneously with multiple parties. It can be a party negotiating in parallel negotiation threads for the same good with different opponent parties \cite{nguyen04,an06,williams12,mansour12} or a party that negotiates simultaneously with multiple parties like in the Contract-Net protocol, and the English and Dutch auction \cite{smith80,sandholm93,shoham09}. Many-to-many negotiations consider the fact that many parties negotiate with many parties, the double auction being the most representative example \cite{shoham09}. Differently to the aforementioned concepts, negotiation teams are not related with the cardinality of the parties but the nature of the party itself. When addressing a negotiation team, we consider a negotiation party that is formed by more than a multiple individuals whose preferences have to be represented in the final agreement. This complex negotiation party can participate in bilateral negotiations, one-to-many negotiations, or many-to-many negotiations. The reason to model this complex negotiation party instead of as multiple individual parties is the potential for cooperation. Despite having possibly different individual preferences, a negotiation team usually exists because there is a shared common goal among team members which is of particular importance.

As far as we are concerned,  only our previous works \cite{sanchez-anguix11,sanchez-anguix12,sanchez-anguix12b,sanchez-anguix13} have considered negotiation teams in computational models. More specifically, the four different computational models introduced in this article are analyzed in different negotiation conditions when facing opponents governed by time tactics.  However, the analysis does not include variability with respect to the strategy carried out by the opponent like the experiments carried out in this present article.

\section{Conclusions and Future Work}
\label{Sec-Conclusion}

This paper presents preliminary results on the performance of existing intra-team strategies for bilateral negotiations against heterogeneous opponents: competitors, conceders, and matchers.  According to our analysis, the intra-team strategies like Full Unanimity Mediated (FUM), Similarity Borda Voting (SBV), Simple Similarity Voting (SSV) and Representative (RE) are able to negotiate with different success against different types of heterogeneous opponents. For the average utility of team members on the final agreement and the joint utility of both parties, we found similar results. In the case of conceders, FUM, SBV, and SSV seems the best options as long as they wait for the opponent to concede and exploit conceders. In the case of matchers, using either FUM or RE employing Agent K's negotiation strategy seem the best choices. This suggests that, for certain types of opponents, a representative approach with an appropriate negotiation strategy may be enough in practice. Finally, the results against competitors show that while strategies like FUM obtain reasonably good results against some competitors like Agent K, all of them suffer from exploitation against other competitor agents like IAMHaggler and IAMCrazyHaggler. Since existing intra-team strategies such as FUM, SBV, and SSV employ time tactics, they are inclined to concede during the negotiation. This may suggest that new intra-team strategies are needed to tackle negotiations against a broader set of competitors.

Additionally, we have extended the well-known negotiation testbed, G\textsc{ENIUS} to support bilateral negotiations where at least one of the parties is a team. The extension allows developers to design their own intra-team strategies by extending the type of mediator used by the team and the type of team member. We expect that by extending G\textsc{ENIUS} with negotiation teams, the research in negotiation teams will further advance and facilitate research.

For future work, we consider designing intra-team negotiation strategies that analyze the behavior of opponent and act accordingly. If a team understands that the opponent is cooperative, the team may act cooperatively and find a mutually acceptable agreement early. Otherwise, if the opponent is a competitor, the team may decide to take a strong position and not concede during the negotiation.

\section{Acknowledgments}
One part of this research is supported by TIN2011-27652-C03-01 and TIN2012-36586-C03-01 of the Spanish government.  Other part of this research is supported by the Dutch Technology Foundation STW, applied science division of NWO and the Technology Program of the Ministry of Economic Affairs; the Pocket Negotiator project with grant number VICI-project 08075 and the New Governance Models for Next Generation Infrastructures project with NGI grant number 04.17. We would also like to thank Tim Baarslag due to his helpful and valuable comments and feedback about G\textsc{enius}.

\bibliography{NTGenius}
\bibliographystyle{abbrv}

\end{document}